\documentclass{PoS}
\usepackage[utf8]{inputenc}

\title{The photon PDF from high-mass Drell Yan data at the LHC}

\ShortTitle{The photon PDF from high-mass Drell Yan data at the LHC}

\author{\speaker{Francesco Giuli}\thanks{On behalf of the analysis team}\\
        University of Oxford\\
        E-mail: \email{francesco.giuli@cern.ch}}

\abstract{
In this contribution, we review the results of~\cite{Giuli}, where a determination of the photon PDF from fits to recent ATLAS measurements of high-mass Drell-Yan dilepton production at $\sqrt{s}$ = 8 TeV is presented.}

\FullConference{XXV International Workshop on Deep-Inelastic Scattering and Related Subjects\\
		3-7 April 2017\\
		University of Birmingham, UK}

\begin{document}

\tableofcontents

\paragraph{Data and theory}
In~\cite{Giuli}, the photon content of the proton, $x\gamma(x,Q^2)$, is extracted from a PDF analysis based on the combined inclusive DIS cross-section data from HERA~\cite{Abramowicz:2015mha} supplemented by the ATLAS measurements of high-mass Drell-Yan (DY) differential cross sections at $\sqrt{s}=8$ TeV~\cite{Aad:2016zzw}. The HERA inclusive data provide information on the quark and gluon content of the proton, while the high-mass DY data provide a direct sensitivity to the photon PDF.
The ATLAS high-mass DY 8 TeV measurements are presented in terms of both the single-differential (1D) invariant-mass distribution, $d\sigma/dm_{ll}$, as well as double\- differential (2D) distributions in $m_{ll}$ and $y_{ll}$, namely  $d^{2}\sigma/dm_{ll}d|y_{ll}|$, and in $m_{ll}$ and $\Delta\eta_{ll}$, $d^{2}\sigma/dm_{ll}\Delta\eta_{ll}$. 
The DIS structure functions and PDF evolution are computed with the {\tt APFEL} program~\cite{Bertone:2013vaa}, which is currently accurate up to NNLO in QCD and NLO in QED, including the relevant mixed QCD+QED corrections. For the calculation of NLO high-mass DY cross sections, the { \tt MadGraph5{\_}aMC@NLO}~\cite{Alwall:2014hca} program is used, which includes the contribution from photon-initiated (PI) diagrams, interfaced to {\tt APPLgrid}~\cite{Carli:2010rw} through {\tt aMCfast}~\cite{amcfast}.
In order to achieve NNLO QCD and NLO EW accuracy in our theoretical calculations, the NLO QCD and LO QED cross sections have been supplemented by bin-by-bin $K$-factors, obtained from {\tt FEWZ}~\cite{Gavin:2012sy} and defined as:
\begin{equation}
  \label{eq:kfactor}
  K(m_{ll},|y_{ll}|) \equiv\frac{\rm NNLO\  QCD  + NLO\  EW}{\rm NLO\  QCD + LO\  EW} \, ,
\end{equation}
using the MMHT2014 NNLO~\cite{Harland-Lang:2014zoa}  PDF set both in the numerator and in the denominator. 
The $K$-factors vary between 0.98 and 1.04, highlighting the fact that higher-order (HO) corrections to the DY process are moderate, in particular at low values of $m_{ll}$ and in the central rapidity region.

\paragraph{Specification of fit inputs}
The input settings of the PDF fits, including the details about the parametrisation of the photon PDF are briefly discussed in the following. First of all, the scale $Q_0^2$ at which PDFs are parametrised is taken to be $Q_0^2 = 7.5~$GeV$^2$, which coincides with the kinematic cut, $Q_{\rm min}^2$, for the data points that are used as input to the fits. The expression for the $\chi^2$  function used for the fits is the one is Ref.~\cite{Aaron:2012qi}, which includes corrections for possible biases from statistical fluctuations and treats the systematic uncertainties multiplicatively. In this analysis, the parametrised PDFs are the valence distributions $xu_{v}(x,Q_0^2)$ and $xd_{v}(x,Q_0^2)$, the gluon distribution $xg(x,Q_0^2)$, and the \textit{u}-type and \textit{d}-type sea-quark distributions, $x\bar{U}(x,Q_0^2)$, $x\bar{D}(x,Q_0^2)$, where $x\bar{U}(x,Q_0^2) = x\bar{u}(x,Q_0^2)$ and $x\bar{D}(x,Q_0^2) = x\bar{d}(x,Q_0^2) + x\bar{s}(x,Q_0^2)$. In addition, the photon distribution $x\gamma(x,Q_0^2)$ is also parametrised at the starting scale.\\
The explicit form of PDF parametrisation at the scale $Q_0^2$ is determined by the technique of saturation of the $\chi^{2}$, namely the number of parameters is increased  one by one  until the $\chi^{2}$ does not improve further, employing Wilks' theorem~\cite{Wilks:1938dza}. Following this method, the optimal parametrisation for the quark and gluon PDFs found for this analysis is 
$xu_v(x) = A_{u_v}x^{B_{u_v}}(1-x)^{C_{u_v}}(1+E_{u_v}x^{2}), xd_v(x) = A_{d_v}x^{B_{d_v}}(1-x)^{C_{d_v}}, x\bar{U}(x) = A_{\bar{U}}x^{B_{\bar{U}}}(1-x)^{C_{\bar{U}}}, x\bar{D}(x) = A_{\bar{D}}x^{B_{\bar{D}}}(1-x)^{C_{\bar{D}}}, xg(x) = A_{g}x^{B_{g}}\\
 (1-x)^{C_{g}}(1+E_{g}x^{2})$, while for the photon PDF it is used
$x\gamma(x) = A_{\gamma}x^{B_{\gamma}}(1-x)^{C_{\gamma}}(1+D_{\gamma}x+E_{\gamma}x^{2})$.
PDF uncertainties are estimated using the Monte Carlo replica method~\cite{DelDebbio:2004xtd,DelDebbio:2007ee,Ball:2008by}, cross-checked with the Hessian method~\cite{Pumplin:2001ct} using $\Delta\chi^2=1$. The former is expected to be more robust than the latter, due to the potential non-Gaussian nature of the photon PDF uncertainties~\cite{Ball:2013hta}.

\begin{figure}[t]
\centering
\includegraphics[width=7.5cm]{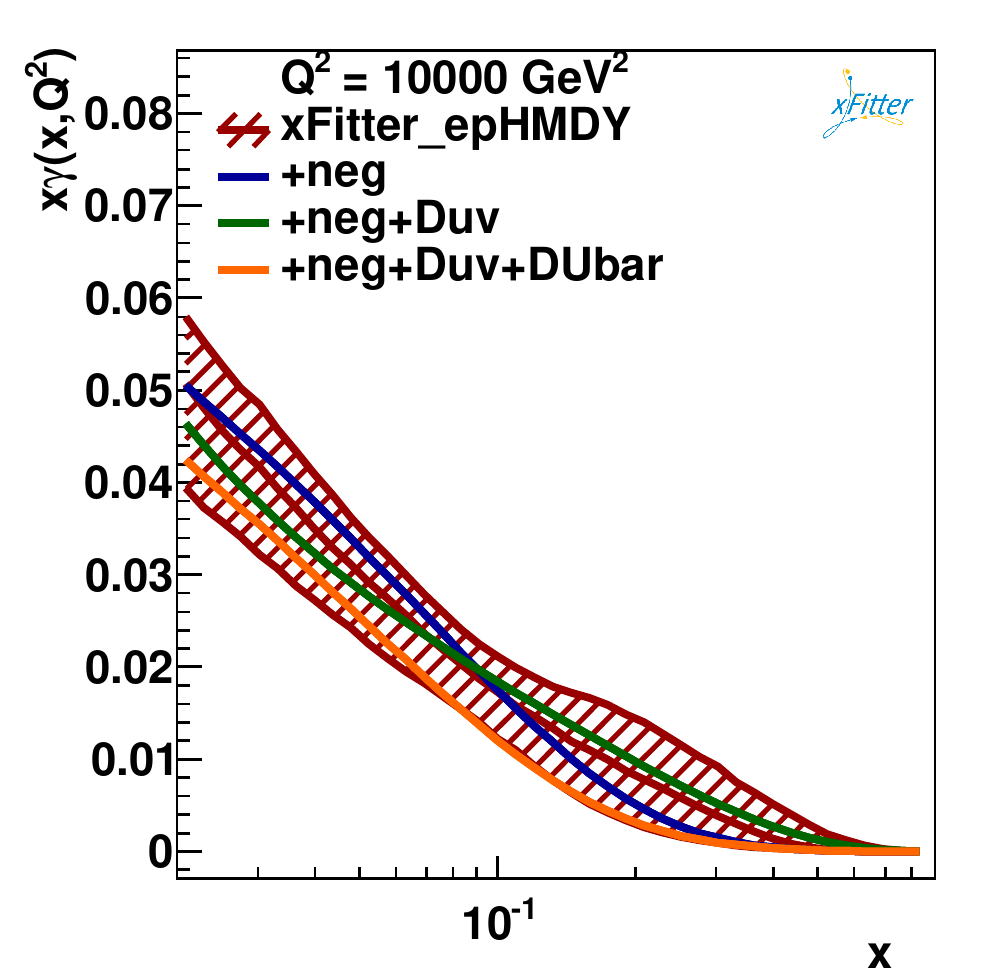}
\includegraphics[width=7.5cm]{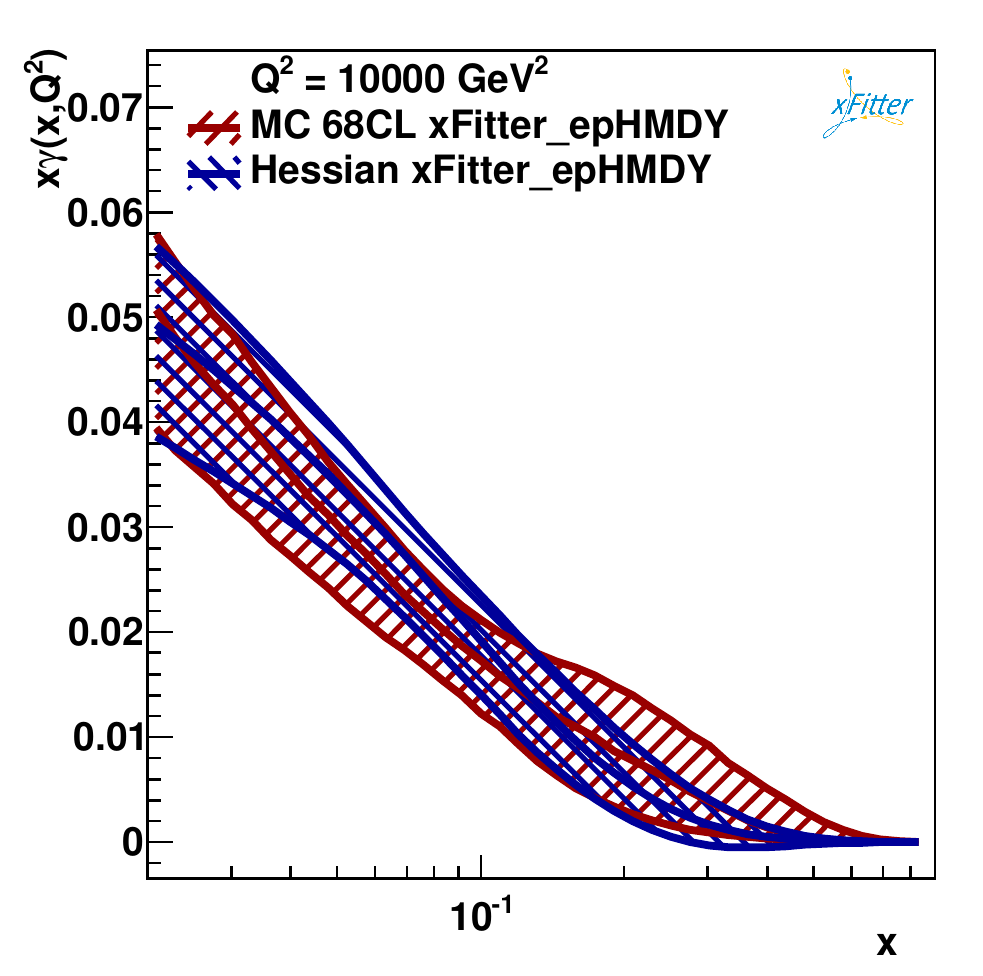} 
\caption{Left: the impact on the photon PDF $x\gamma(x,Q^2)$ from {\tt xFitter\_epHMDY} in fits where a number of additional free parameters are allowed in the PDF parametrisation. Right: comparison between the {\tt xFitter\_epHMDY} determinations obtained with the Monte Carlo (baseline) and with the Hessian methods, where in both cases the PDF error band  shown corresponds to the 68\% CL uncertainties. }
\label{fig:param}
\label{fig:photon_mc_vs_hessian}
\end{figure}

\paragraph{Results}
\begin{figure}[t]
\centering
\includegraphics[width=7.5cm]{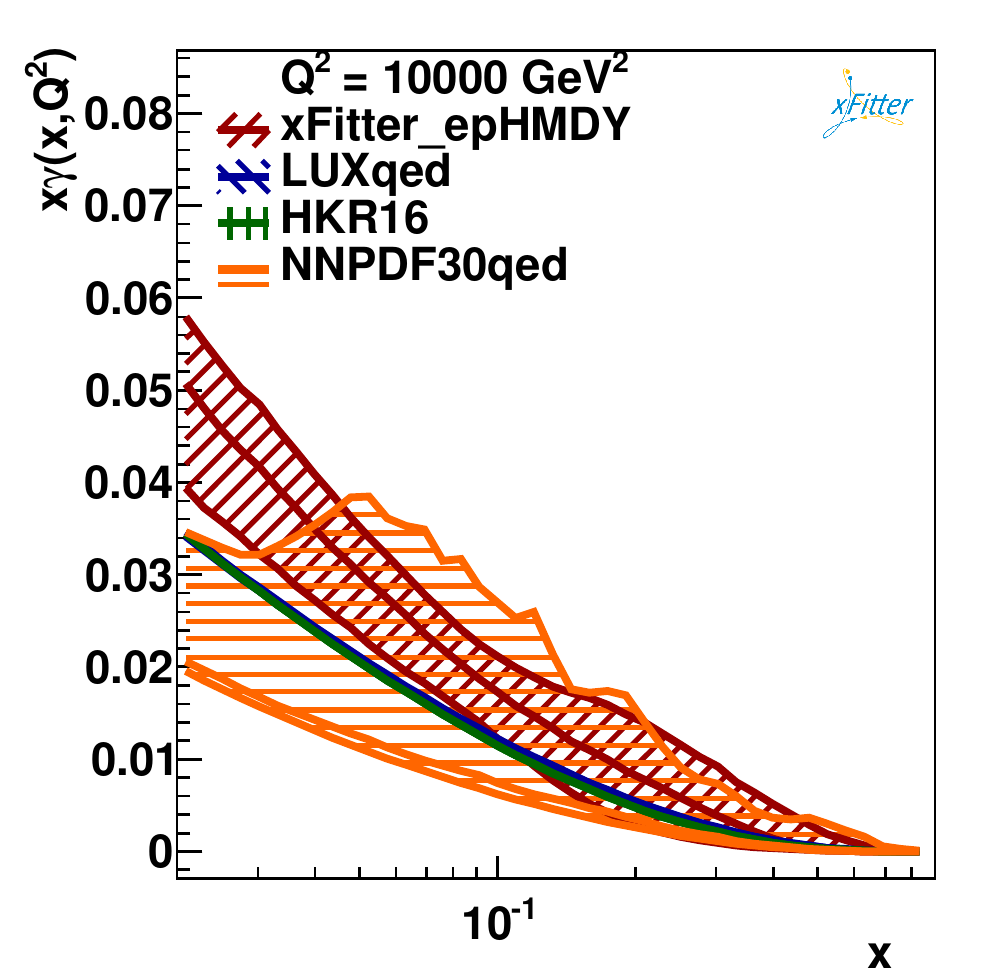}
\includegraphics[width=7.5cm]{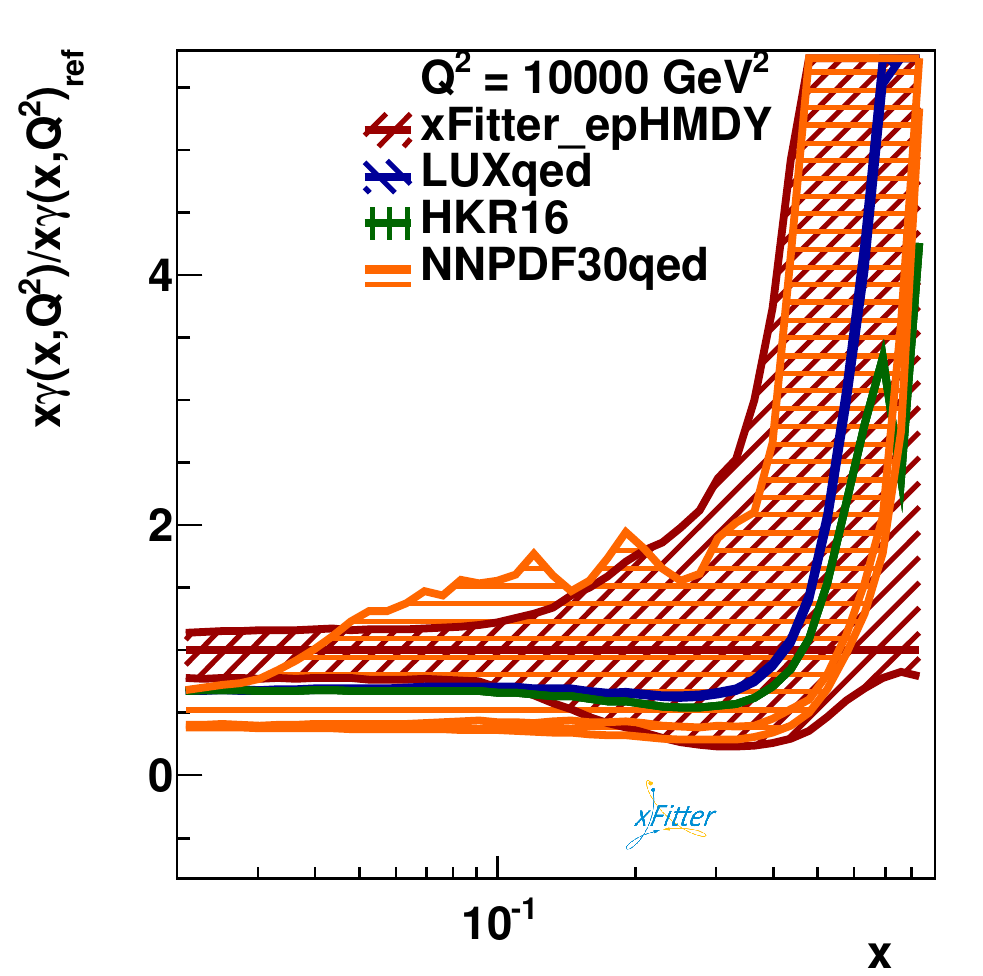} 
\caption{Left: Comparison between the photon $x\gamma(x,Q^2)$ at $Q^2=10^4$ GeV$^2$ from the present NNLO analysis ({\tt xFitter\_epHMDY}) with the corresponding results from NNPDF3.0QED, LUXqed and HKR16. Right: the same comparison, now with the results normalized to the central value of {\tt xFitter\_epHMDY}. For the present fit, the PDF uncertainties are shown at the 68\% CL obtained from the MC method. For HKR16 only the central value is shown, while for LUXqed the associated PDF uncertainty band is included. }
\label{photon_zoom} \label{photon_zoom_ratio}
\end{figure}
\begin{figure}[t]
\centering
\includegraphics[width=7.5cm]{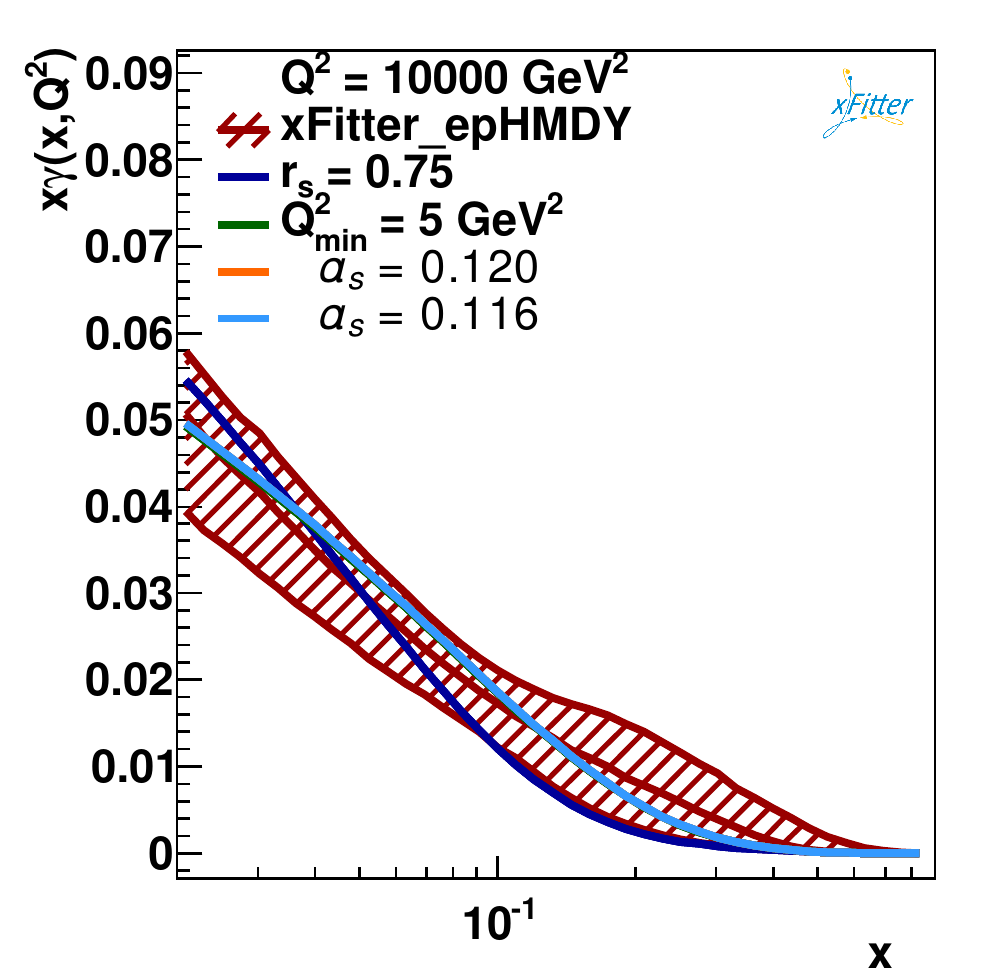}
\includegraphics[width=7.5cm]{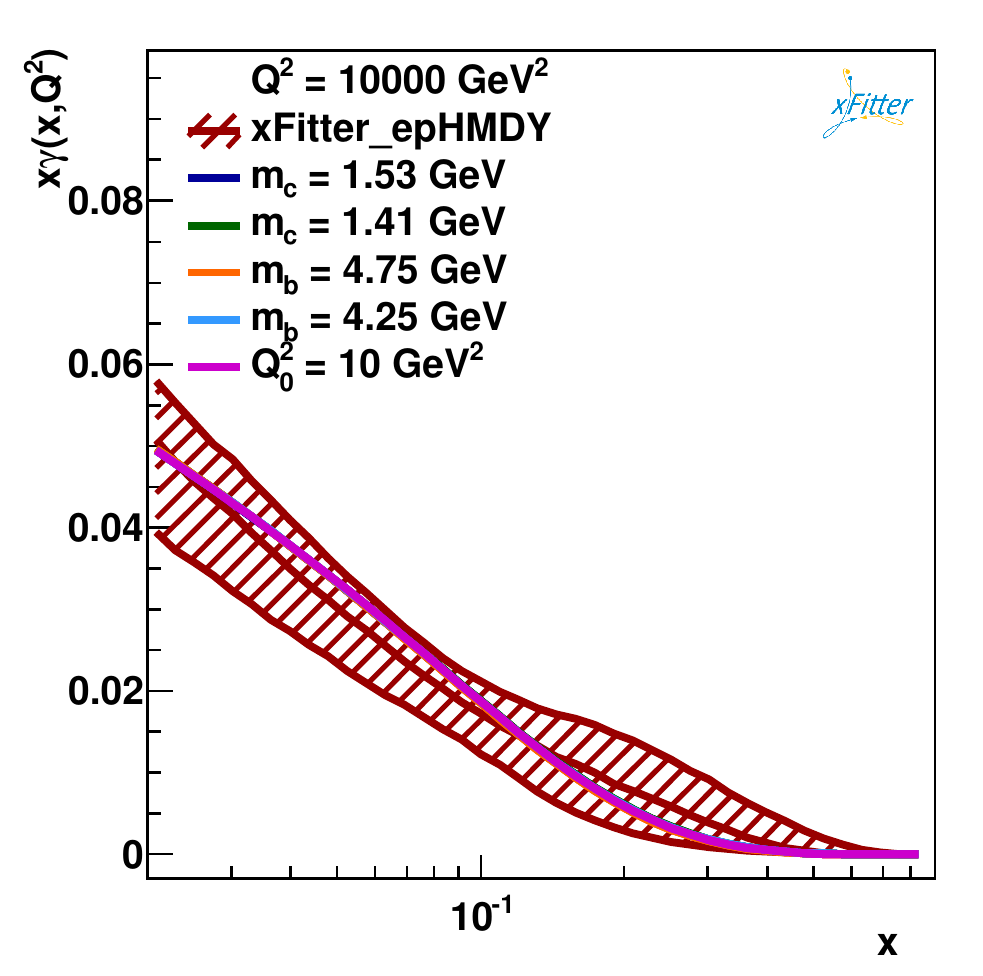}
\caption{Comparison between the baseline determination of $x\gamma(x,Q^2)$ at $Q^2=10^4$ GeV$^2$ in the present analysis, {\tt xFitter\_epHMDY}, with the central value of a number of fits for which one input parameter has been varied. The curves are indistinguishable because they overlap due to their negligible impact on photon PDF fit}
\label{fig:model}
\end{figure}
In the following, the results that will be shown correspond to those obtained from fitting the double-differential $(m_{ll},y_{ll})$ cross-section distributions. It has been cross-checked that comparable results are obtained if the $(m_{ll},\Delta\eta_{ll})$ cross-section distributions are fitted instead. For the baseline NNLO fit, the value $\chi^2_{min}/N_{dof} = 1284./1083$ is obtained where $N_{dof}$ is the number of degrees of freedom in the fit which is equal to total number of data points minus number of free parameters. The contribution from the HERA inclusive  data is  $\chi^{2}/N_{\rm dat} = 1236/1056$ and from the ATLAS high-mass DY data is $\chi^2/N_{\rm dat} = 48/48$, where  $N_{\rm dat}$ the number of the data points for the corresponding data sample.\\ 
In Fig.~\ref{photon_zoom}, the photon PDF, $x\gamma(x,Q^2)$, is shown at $Q^2=10^4$ GeV$^2$,  and it is compared to the corresponding NNPDF3.0QED~\cite{Ball:2013hta}, LUXqed~\cite{Manohar:2016nzj} and  HKR16~\cite{hkr2016} results. In the left plot, the comparison is shown in an absolute scale, while the plot on right shows the ratio of different results normalized to the central value of the fit. For the present fit, {\tt xFitter\_epHMDY},  the experimental PDF uncertainties at the 68\% confidence level (CL) are obtained using the Monte Carlo (MC) method. In Fig.~\ref{photon_zoom}, the $x$-range is set between 0.02 and 0.9, since outside that region there is only limited sensitivity to the photon PDF. Fig.~\ref{photon_zoom} shows that the four determinations of the photon PDF are consistent within PDF uncertainties for $x\ge 0.1$, while, for smaller values of $x$, the photon PDF from LUXqed and HKR16 is smaller than {\tt xFitter\_epHMDY}, but still in agreement at the 2-$\sigma$ level. Fig.~\ref{photon_zoom} shows that for $0.04 \le x \le 0.2$ the present analysis  exhibits smaller PDF uncertainties as compared to those from  NNPDF3.0QED.\\
The impact of the high-mass Drell-Yan 8 TeV measurements on the light quark PDFs has been assessed and, as expected, while these data have a significant constraint on the photon PDF, their impact on the quark and gluon PDFs is rather moderate.\\
Moreover, a number of variations has been assessed in order to test the robustness of this $x\gamma(x,Q^2)$ determination.First of all, Fig.~\ref{fig:model} shows the comparison between the {\tt xFitter\_epHMDY} determination of photon PDF at $Q^2=10^4$ GeV$^2$, including the experimental MC uncertainties, with the central value of those fits for which a number of variations have been performed. Specifically: the strong coupling constant is varied by $\delta \alpha_s=\pm ~0.002$ around the central value; the ratio of strange to non-strange light quark PDFs is decreased to $r_s=0.75$ instead of $r_s=1$ (nominal value in the fits); the value of the charm mass is varied between $m_c=1.41$ GeV and $m_c=1.53$ GeV, and that of the bottom mass between $m_b=4.25$ GeV and $m_b=4.75$ GeV; the minimum value $Q_{\rm min}^2$ of the fitted data is decreased down to $5$ GeV$^2$; the input parametrisation scale $Q_0^2$ is raised to $10$ GeV$^2$ as compared to the baseline value of $Q_0^2=7.5~$GeV$^2$. The results shown in Fig.~\ref{fig:model} highlight that in all cases, the effect of the variations considered here is contained within (and typically much smaller than) the experimental PDF uncertainty bands of the reference fit. In each case, one variation at a time is performed and compared with the central value of $x\gamma(x,Q^2)$ and its experimental PDF uncertainties computed using the MC method.\\
Then, the $x\gamma(x, Q^2)$ determination is compared with further fits where a number of new free parameters are allowed in the PDF parametrisation. Fig.~\ref{fig:param} shows the impact of three representative variations: more flexibility to the gluon parametrisation by adding a negative term (labelled by "neg" - note that the parametrisation does not actually become negative within the fitted $x,Q^{2}$ range of the data); secondly, a parameter $D_{u_{v}}$ is added to the $u_{v}$ distribution, making its multiplying polynomial (1+$D_{u_{v}}x$+$E_{u_{v}}x^{2}$); thirdly, $D_{\bar{U}}$ is also added, making its multiplying polynomial (1+$D_{\bar{U}}x$). As before, all variations are contained within the experimental PDF uncertainty bands, though the impact of the parametrisation variations is typically larger than that of the model variations: in the case of the ${\rm neg}+D_{u_{v}}+D_{\bar{U}}$ variations, the central value is at the lower edge of the PDF uncertainty band in the entire range of $x$ shown.\\
Furthermore, the robustness of the estimated experimental uncertainty of the photon PDF in this analysis has been estimated, providing the comparison between the MC and Hessian methods. This comparison is presented in the right plot in Fig.~\ref{fig:photon_mc_vs_hessian} and it indicates a reasonable agreement between the two methods. In particular, the central values of the photon obtained with these two different fitting techniques are quite similar to each other. As expected, the MC uncertainties tend to be larger than the Hessian ones, specially in the region $0.2\lesssim x$, indicating deviations with respect to the Gaussian behaviour of the photon PDF.\\
Finally, the perturbative stability of the {\tt xFitter\_epHMDY} determination of the photon PDF with respect to the inclusion of higher order QCD corrections in the analysis has been quantified. A comparison between the baseline fit of $x\gamma(x,Q^2)$, based on NNLO QCD and NLO QED theoretical calculations, with the central value resulting from a corresponding fit based instead on NLO QCD and QED theory has been made (in other words, the QED part of the calculations is identical in both cases). The fit of $x\gamma(x,Q^2)$ exhibits a reasonable perturbative stability, since the central value of the NLO fit is always contained in the 1-sigma PDF uncertainty band of the baseline {\tt xFitter\_epHMDY} fit. The agreement between the two fits is particularly good for $0.1\lesssim x$, where the two central values are very close to each other. This comparison is shown at low scale, $Q^2=7.5$ GeV$^2$, and higher scale, $Q^2=10^4$ GeV$^2$, indicating that perturbative stability is not scale dependent.

\paragraph{Conclusions}
In this contribution, we reviewed the results presented in~\cite{Giuli}, where a new determination of the photon PDF from a fit of HERA inclusive DIS structure functions supplemented by ATLAS data on high-mass DY cross sections has been performed, based on the {\tt xFitter} framework.The results of this study exhibit smaller PDF uncertainties that the only other existing photon PDF fit from LHC data, the NNPDF3.0QED analysis, based on previous LHC DY measurements. These new results are in agreement within uncertainties with two recent theoretical calculations of the photon PDF, LUXqed and HKR16. For $x\ge 0.1$, the agreement is at the 1-$\sigma$ level already including only the experimental MC uncertainties, while for $0.02 \ge x \ge 0.1$ it is important to account for parametrisation uncertainties. The findings indicate that a direct determination of the photon PDF from hadron collider data is still far from being competitive with the LUXqed and HKR calculations, which are based instead on precise measurements of the inclusive DIS structure functions of the proton.

\end{document}